\documentclass{article}
\input{tcilatex}

\begin{document}

\title{Two-Body Dirac Equations from Relativistic Constraint Dynamics with
Applications to QED and QCD Bound-States and to N-N Scattering}
\author{H. \textsc{Crater}, B. \textsc{Liu} \\
%EndAName
The University of Tennessee Space Institute, Tullahoma, TN \and P. \textsc{%
Van Alstine} \\
%EndAName
12474 Sunny Glenn Dr. Moorpark, Ca.}
\maketitle

\section{Introduction}

The formulation of relativistic two-body bound state wave equations and
their relationship to quantum field theory began with the work by Eddington
and Gaunt in 1928 \cite{edgnt}. However, the large variety of approaches
attempted in recent years shows that this problem still has no generally
agreed-upon solution. Perhaps for this reason, most recent field theory
books have skirted this topic. In his recent text, Steven Weinberg states \ 
\cite{wnbrg}: \textquotedblleft It must be said that the theory of
relativistic effects and radiative corrections in bound states is not yet in
satisfactory shape.\textquotedblright

Of course, this topic is often presented as covered by \ the manifestly
covariant Bethe-Salpeter equation obtained directly from relativistic
quantum field theory. Over the years, however, many problems have turned up
to impede its direct implementation, mostly related to the central role
played in it by the relative time or energy\cite{nak}. These difficulties
have led many authors to attempt reformulations.

We describe here a recent approach resulting from \textquotedblright\
Two-Body Dirac Equations\textquotedblright\ (emerging from Dirac's
Relativistic Constraint Dynamics) that does satisfy many of the requirements
one would demand of a treatment of the relativistic two-body problem. We use
its applications to QED bound states such as positronium, QCD quarkonia, and
the nucleon-nucleon scattering problem to demonstrate the advantages of the
approach.

\section{The One-Body Dirac Equation}

For a single body, the original Dirac equation for a single spin-one-half
particle (nearly universally accepted) provides a successful bound state
equation. \ The free Dirac equation $(\gamma \cdot p+m)\psi =0$ serves as
a~relativistic~version~of~Newton's~1st~law.$~$When the four-vector
substitution for electromagnetic interaction $p_{\mu }\rightarrow p_{\mu
}-A_{\mu },~$and the minimal mass substitution for scalar interaction$%
~~m\rightarrow m+S$ \ are performed on it, one
obtains~a~relativistic~version~of~Newton's~2nd~law 
\begin{equation}
(\gamma \cdot (p-A)+m+S)\psi =0.
\end{equation}
The two-body Dirac equations of constraint dynamics successfully extend this
one-body minimal coupling form to the interacting two-body system.

\section{Two-Body Dirac Equations from Constraint Dynamics}

In the 1970's, Todorov, Kalb and Van Alstine, and Komar independently used
Dirac's constraint mechanics to attack the relativistic two-body problem for
spinless particles \cite{dirac}. By covariantly controlling the relative
time variable these authors eliminated negative norm states as well as
circumvented the no-interaction theorem of Currie, Jordan, and Sudarshan\cite%
{sudar} that had discouraged further work in this area since the 1960's .\ \
By combining constraint dynamics with particle supersymmetries, Crater and
Van Alstine extended those works to pairs of spin one half particles to
obtain two-body quantum bound state equations that correct not only defects
in the Breit equation but those in the ladder approximation to the
Bethe-Salpeter equation as well. \cite{van}These Two-Body Dirac Equations of
constraint dynamics possess a number of important features (some of which
are unique) which provide an alternative formulation of fundamental
field-theoretic results (while yielding standard perturbative spectra) and
correct defects in phenomenological applications that result from patchwork
introduction of interactions : They a) provide a three dimensional but
covariant rearrangement of the Bethe-Salpeter equation, b) yield simple
three-dimensional Schr\"{o}dinger-like forms similar to their
nonrelativistic counterparts, c) contain spin dependences determined
naturally by their incorporation of Dirac's one-body structures, d) contain
well defined strong potential structures that pass the necessary test that
they reproduce correct QED perturbative results when solved
nonperturbatively, e) in phenomenological applications make unnecessary the
ad hoc introduction of cutoff parameters generally used to avoid singular
potentials and f) have relativistic potentials which may be related directly
to the interactions of perturbative quantum field theory or (e.g. for QCD)
may be introduced semiphenomenologically\cite{recent}.These equations
provide a non-perturbative or strong-potential framework for extrapolating
perturbative field theoretic results into the highly relativistic regime of
bound light particles in a quantum mechanically well defined way.

\subsection{ World Vector and Scalar Interactions}

The constraint formalism is embodied in a system of two coupled, compatible
Dirac equations on a single wave-function. For particles interacting through
world vector and scalar interactions the two-body Dirac equations take the
general minimal-coupling form 
\begin{eqnarray}
\mathcal{S}_{1}\psi &\equiv &\gamma _{51}(\gamma _{1}\cdot (p_{1}-\tilde{A}%
_{1})+m_{1}+\tilde{S}_{1})\psi =0  \nonumber \\
\mathcal{S}_{2}\psi &\equiv &\gamma _{52}(\gamma _{2}\cdot (p_{2}-\tilde{A}%
_{2})+m_{2}+\tilde{S}_{2})\psi =0.  \label{tbde}
\end{eqnarray}%
The two equations are compatible in the sense that, 
\begin{equation}
\lbrack \mathcal{S}_{1},\mathcal{S}_{2}]\psi =0.~~~~~
\end{equation}%
\ This condition is satisfied as a result of the presence in these equations
of spin supersymmetries, a relativistic 3rd law, and covariant restrictions
on the relative time and energy. Its direct dynamical consequence\ is the
automatic incorporation of correct spin-dependent recoil terms .

For the case where only vector interactions are present, the compatibility
condition is most naturally satisfied by vector potentials in the
\textquotedblleft hyperbolic\textquotedblright\ momentum and spin-dependent
forms

\begin{equation}
\tilde{A}_{1}=[1-\text{$\cosh $}(\mathcal{G})]p_{1}+\sinh (\mathcal{G})p_{2}-%
\frac{i}{2}(\partial \exp \mathcal{G}\cdot \gamma _{2})\gamma _{2}
\end{equation}
\begin{equation}
\tilde{A}_{2}=[1-\cosh (\mathcal{G})]p_{2}+\sinh (\mathcal{G})p_{1}+\frac{i}{%
2}(\partial \exp \mathcal{G}\cdot \gamma _{1})\gamma _{1}.
\end{equation}
$\ $ In this case, the compatibility condition enforces a ''third law'' with
the two constituent potentials actually depending on only one invariant $%
\mathcal{A}.$through the interaction function $\mathcal{G(A)}.~\ $The
explicit form of $\mathcal{G(A)}$ follows both from comparison with
classical Wheeler-Feynman electrodynamics or with QED through leading term
summation of ladder, cross-ladder and constraint diagrams \cite{whlr} 
\begin{equation}
\mathcal{G(A)=}-\frac{1}{2}\log (1-\frac{2\mathcal{A}}{w})~;~~w~\text{the
total c.m. energy~.}
\end{equation}

Compatibility also requires that the relative time be covariantly controlled
through interactions depending only on $x_{\perp }$, a covariant spacelike
particle separation variable$\mathrm{~}$perpendicular~to the total momentum$%
~P$ 
\begin{eqnarray}
\mathcal{A} &&\mathcal{=}\mathcal{A(}x_{\perp })  \nonumber \\
x_{\perp }^{\mu } &=&x^{\mu }+\hat{P}^{\mu }(\hat{P}\cdot x),~~~\hat{P}%
\equiv \frac{P}{w}~\text{is a time-like unit vector}.
\end{eqnarray}
For lowest order electrodynamics,

\begin{equation}
\mathcal{A}=\mathcal{A}(x_{\perp })=-\frac{\alpha }{r}\ \mathrm{;}r\equiv 
\sqrt{x_{\perp }^{2}}.
\end{equation}

For quark-models, we must include scalar potentials $\tilde{S}_{i}$ . \ When
appearing with vector interactions they depend not only on two invariant
mass potential functions $M_{1}(x_{\perp }),M_{2}(x_{\perp }),$ related to
each other through one invariant function $L(x_{\perp })$ but also on the
vector interaction through $\mathcal{G(A(}x_{\perp }))$ 
\begin{eqnarray}
\tilde{S}_{1} &=&M_{1}-m_{1}-\frac{i}{2}\exp \mathcal{G(A)}\gamma _{2}\cdot 
\frac{\partial M_{1}}{M_{2}},  \nonumber \\
\tilde{S}_{2} &=&M_{2}-m_{2}+\frac{i}{2}\exp \mathcal{G(A)}\gamma _{1}\cdot 
\frac{\partial M_{2}}{M_{1}},
\end{eqnarray}%
\begin{equation}
M_{1}^{2}-M_{2}^{2}=m_{1}^{2}-m_{2}^{2}\Longrightarrow 
\begin{array}{c}
M_{1}=m_{1}\ \cosh L\ +m_{2}\sinh L\  \\ 
M_{2}=m_{2}\ \cosh L\ +m_{1}\ \sinh L%
\end{array}%
;3RD~~LAW.  \label{ml}
\end{equation}%
The counterpart to the invariant $\mathcal{A}$ for scalar interactions is $S$
with the form of $L=L(S(x_{\perp }),\mathcal{A(}x_{\perp }\mathcal{))}$ with 
$m_{w}=m_{1}m_{2}/w$ from 
\begin{equation}
M_{i}^{2}=m_{i}^{2}+\exp \mathcal{G(A)}(2m_{w}S+S^{2})~;i=1,2.  \label{ams}
\end{equation}%
Retardative effects are already included through the c.m. energy dependences
of the potential structures. Although the potential forms in these equations
may seem unfamiliar, expansion of the resulting classical dynamics in $1/c$
around the nonrelativistic limit shows that it is canonically equivalent to
order $1/c^{2}$ to the dynamics generated by the corresponding
single-quantum exchange in field theory.

\subsection{Manifest Covariance and Well-Defined Quantum-Mechanical Behavior}

\ For our applications we use a Pauli reduction to bring our equations to
the covariant Schr\"{o}dinger-like form (with $p$ the relative momentum) 
\begin{equation}
{\Large (}p^{2}+\Phi _{w}(\sigma _{1},\sigma _{2},p_{\perp },\mathcal{A}%
(r),S(r)){\Large )}\psi =b^{2}(w)\psi
\end{equation}%
incorporating exact two-body relativistic kinematics through the eigenvalue $%
b^{2}(w)=(w^{4}-2(m_{1}^{2}+m_{2}^{2})w^{2}+(m_{1}^{2}-m_{2}^{2})^{2})/4w^{2} 
$ $\equiv \varepsilon _{w}^{2}-m_{w}^{2}$ in terms of $m_{w}$ and $%
\varepsilon _{w}=(w^{2}-m_{1}^{2}-m_{2}^{2})/2w$ respectively the mass and
energy of the fictitious particle of relative motion. This Schr\"{o}%
dinger-like equation is not only manifestly covariant but quantum
mechanically well defined: one can solve it nonperturbatively in both QED
and QCD bound state cases for which every term in the quasipotential $\Phi
_{w}(\sigma _{1},\sigma _{2},p_{\perp },\mathcal{A}(r),S(r))$ is less
singular than $-1/4r^{2}$(in contrast to all reductions of the Breit
equation and many reductions of the Bethe-Salpeter equation). It involves at
most 2 coupled wave equations but all portions of the 16 component wave
function play essential roles in spectral calculations, either directly or
through the strong potential structures that they generate when they are
eliminated. \ The explicit forms of the spin dependent potentials that
appear in the Pauli-form quasipotential $\Phi _{w}$ are dictated by the
interaction structure of the original two coupled Dirac equations and are
not put in by hand.

The Schr\"{o}dinger-like form of the Two-Body Dirac equations takes the
minimal coupling form 
\begin{equation}
(p^{2}+(m_{w}+S)^{2}-(\varepsilon _{w}-\mathcal{A)}^{2}+\Phi _{sp}(\sigma
_{1},\sigma _{2},p_{\perp },\mathcal{A}(r),S(r)){\Large )}\psi =0
\end{equation}
in which 
\begin{eqnarray}
\Phi _{sp} &=&\Phi _{D1}\hat{r}\cdot p+\Phi _{D2}+\Phi _{SO}L\cdot (\sigma
_{1}+\sigma _{2})+\Phi _{SOD}L\cdot (\sigma _{1}-\sigma _{2})+\Phi
_{SPO}L\cdot (\sigma _{1}\times \sigma _{2})  \nonumber \\
&&+\Phi _{SS}\sigma _{1}\cdot \sigma _{2}+\Phi _{T}\sigma _{1}\cdot \hat{r}%
\sigma _{2}\cdot \hat{r}+\Phi _{DT}\hat{r}\cdot p\sigma _{1}\cdot \hat{r}%
\sigma _{2}\cdot \hat{r}.
\end{eqnarray}
\ The $\Phi _{i}=\Phi _{i}(\mathcal{A},S,w)$ are not independent but are all
determined in terms of $\mathcal{A},S$ $\ $through the Pauli reduction.$~$

We have checked analytically and numerically that our strong potential terms
do not lead to spurious results by solving them nonperturbatively to obtain
agreement with the standard fine and hyperfine spectra of perturbative QED.\
\ For example, for the singlet positronium system with $\mathcal{A}=-\alpha
/r$ \ our fully coupled system of 16-component equations $\mathcal{S}%
_{1}\psi =\mathcal{S}_{2}\psi =0$ is exactly solvable with total energy\cite%
{exct} 
\begin{equation}
w=m(2+2/(1+\alpha ^{2}/(n+((l+1/2)^{2}-\alpha
^{2})^{1/2}-l-1/2)^{2})^{1/2})^{1/2}\dot{=}m(2-\frac{\alpha ^{2}}{4}-\frac{%
21\alpha ^{4}}{64})_{\text{ground state}}
\end{equation}%
Such validation should be required of all candidate equations for
nonperturbative quark model calculations and other semiphenomenological
applications when their quark-model kernals are replaced by ones appropriate
for QED. Otherwise, how can one trust the short distance spectral
contributions obtained when applied in QCD? \ No other approaches have yet
fully passed this test. \ In fact Sommerer et al \cite{vary} have shown that
all of the well-known quasipotential approaches (e.g. the Blankenbecler-
Sugar equation, the Gross, and Kadeshevsky approaches) fail this crucial
test even for the ground state. \ However, by varying parameters permitted
by the non-uniqueness of the quasipotential approaches, Sommerer et al
obtained a quasipotential model that does reproduce this ground state
numerically. We have successfully extended the check in our equations and
found numerical agreement with the perturbative results for a range of
angular momentum and radial states and for unequal masses.\cite{exct}.

The invariant forms for $\mathcal{A}$ $\ $or $S$ \ in our equations follow
both from relativistic classical field theory and from the perturbative
treatment of corresponding quantum field theories\cite{exct} through
Sazdjian's derivation of the constraint equations as a ``quantum mechanical
transform of the Bethe-Salpeter equation''\cite{saz}.

\section{Two-Body Dirac Equations in Meson Spectroscopy}

\subsection{The Adler-Piran Potential}

We obtain a constraint version of the naive quark model for mesons from a
covariant adaptation of a static quark potential due to Adler and Piran.
These authors use an effective non-linear field theory derived from QCD to
find\cite{adl} 
\begin{equation}
V_{AP}(r)=\Lambda (U(\Lambda r)+U_{0})\ (=\mathcal{A}+S).
\end{equation}%
The original $V_{AP}$ is nonrelativistic, appearing as the sum of world
vector and scalar potentials in the nonrelativistic limit. \ $V_{AP}$
incorporates a running coupling constant form in coordinate space at short
distance $\Lambda U(\Lambda r<<1)\sim 1/rln\Lambda r$ and includes linear
confinement plus subdominant logarithm terms at long distance 
\begin{equation}
V_{AP}(r)=\Lambda (c_{1}\Lambda r+c_{2}\log (\Lambda r)+\frac{c_{3}}{\sqrt{%
\Lambda r}}+\frac{c_{4}}{\Lambda r}+c_{5}),\ \ \Lambda r>2.
\end{equation}

When the nonrelativistic quark model is constructed with realistic
potentials such as \ the Adler-Piran potential or Richardson potential it
fails for light mesons. Beyond a certain limit, it gives meson masses that
increase with decreasing quark mass. \ However, this effect is entirely
absent from the light meson spectra produced by the Two-Body Dirac Equations
with their relativistic kinematics and QCD-determined relativistic
potentials..

\subsection{Relativistic Naive Quark Model}

We reinterpret the static $V_{AP}$ covariantly by a) replacing the
nonrelativistic $r$ by $\sqrt{x_{\perp }^{2}}$ and b) parcelling out the
static potential $V_{AP}$ into the invariant functions $\mathcal{A}(r)$ and $%
S(r)$ as follows

\begin{equation}
\mathcal{A}=exp(-\beta r)[V_{AP}-\frac{c_{4}}{r}]+\frac{c_{4}}{r}+\frac{%
e_{1}e_{2}}{r},~~S=V_{AP}+\frac{e_{1}e_{2}}{r}-\mathcal{A}.
\end{equation}%
This partially phenomenological step ensures that at short distance the
potential is strictly vector while at long distance the vector portion is
strictly Coulombic with the confining portion at long distance (including
subdominant portions) strictly scalar. Note that in our equations, once $%
\mathcal{A}$ and $S$ have been determined, so are all the accompanying
spin-dependent interactions. \ Here our approach is that of a naive quark
model since we ignore flavor mixing and the effects of decays on the bound
state energies.

\subsection{Meson Spectroscopy}

Our results are spectrally quite accurate, from the heaviest upsilonium
states to the pion. Notable exceptions are light meson orbital and radial
excitations and their spin-orbit splittings. \ We include only selected
portions of the whole table here (see \ hep-ph-0208186 for the complete
results). The parameter values we obtain from the best fit are $%
m_{b}=4.877,~m_{c}=1.507,~m_{s}=0.253,~m_{u}=0.0547,~m_{d}=0.0580$ GeV along
with $\Lambda =0.216,~\Lambda U_{0}=1.865$ GeV and $\beta =1.936$. \ We find
that the static Adler-Piran potential, having a close connection with
quantum chromodynamics (QCD), gives a good fit to the mostly nonrelativistic
bottomonium spectrum. We obtain 9.453, 9.842, 9.889,9.921,10.022 GeV for the
ground state and first orbital and radial states of
9.460,9.860,9.892,9.913,10.023 GeV. For the B mesons the fit results from
the fact that for them our equations essentially reduce to the one-body
Dirac equation \ We obtain for the $B(5.279),B^{\ast
}(5.325),B_{s}(5.369),B_{s}^{\ast }(5.416)$ mesons the results
5.273,5.321,5.368,5.427 GeV. \ The goodness of the fit to the charmed quark
mesons shows that the equations perform well in the semirelativistic region.
The ground state ($^{1}S_{0},^{3}S_{1})$ and first orbital ($%
^{1}P_{1},^{3}P_{0},^{3}P_{1},^{3}P_{2})$ and radial (2$%
^{1}S_{0},2^{3}S_{1}) $ excitations of 2.980, 3.097, 3.526,
3.415,3.510,3.556,3.594,3.686 GeV match our theoretical values of 2.978,
3.129,3.520, 5.407,3.507, 3.549,3.610,3.688 GeV quite nicely. The same thing
happens for the $D(1.865), $ $D^{\ast }(2.007),$ $D_{s}(1.968),D_{s}^{\ast
}(2.112)$ mesons for which we obtain the values of 1.866,2.000,1.976,2.123
GeV. The goodness of the accompanying fit to the lighter mesons (with the
same two invariant potential functions used for the entire spectrum) is due
to exact two-body relativistic kinematics combined with the minimal
interaction and strong potential structures of our equations for vector and
scalar potentials. \ For example, for the ground states $K(494),K^{\ast
}(892),\phi (1.019),\pi (140),$and $\rho (767)$, we obtain
0.492,0.910,1.033, 0.144, 0.792 GeV. The results for the light meson orbital
and radial excitations and \ their spin-orbit splittings are mixed. \ For
example the $b_{1}(1.231),a_{0}(1.450),a_{1}(1.230),a_{2}(1.318)$ meson fits
of 1.392,1.491,1.568,1.310 are quite uneven and the radially excited $\pi
(1.300)$ and $\rho (1.465)$ results of 1.536 and 1.775 GeV are quite far off
the mark. \ On the other hand the results 1.319,1.533,1.493 GeV for the $%
\phi $ orbital excitations are reasonable. In the future, we plan to use a
coupled channel formalism to investigate the origin of some of these
problems.

The strength of the Two-Body Dirac approach is that using it , with just two
parametric functions $\mathcal{A}$ and $S$ we are able to obtain an overall
fit about as good as that obtained by Godfrey and Isgur\cite{isg}, who used
six parametric functions, basically one for each type of spin dependence.

As a bonus, we find that the pion is a Goldstone boson in the sense that $%
m_{\pi }(m_{q}\rightarrow 0)\rightarrow 0$ while the $\rho $ and excited $%
\pi $ have finite mass in this limit (see hep-ph/0208186 for complete
plots). We have shown elsewhere that $\langle 0|\partial _{\mu }j_{5}^{\mu
}|\pi \rangle =(M_{1}(0)+M_{2}(0))=(m_{1}+m_{2})Tr(\gamma _{5}\psi )$\
supporting our claim that our potential model incorporates aspects of a
spontaneous breakdown of chiral symmetry.

\section{ Two-Body Dirac Equations in Nucleon-Nucleon Scattering}

\ The two-body Dirac equations of constraint dynamics provide a natural
method for extending nonrelativistic phenomenological treatments to the
relativistic domain with effective potentials determined from the standard
field-theoretic treatment of meson exchanges. This provides a severe test
for the strong potential terms that turned out to be essential for the
treatment of \ QED and QCD bound states. The mesons we include are the
pseudoscalar mesons $\pi (135),\eta (548),\eta ^{\prime }(952)$ the vector
mesons $\rho (770),\omega (776),\phi (1020)$ and the scalar mesons $\sigma
(600),a_{0}(980),f_{0}(983).$ \ The $\pi ,\rho ,$ and the $a_{0}$ are
isovector mesons while the rest are isoscalar mesons. \ To incorporate these
nine meson exchange forces into the two-body Dirac equations we need to
generalize the interactions contained in our equations to include others
beyond world-scalar and world-vector.

\subsection{Two-Body Dirac Equations for General Covariant Interactions: The
role of supersymmetry}

The detailed forms of interaction in our equations are actually the
consequences of supersymmetries in our interacting two-body system. As an
illustration, we review the derivation of Eq.(\ref{tbde}). Define theta
matrices in terms of Dirac matrices $\theta ^{\mu }\equiv i\sqrt{1/2}\gamma
_{5}\gamma _{,}^{\mu }\ \mu =0,1,2,3,\ \ \theta _{5}\equiv i\sqrt{1/2}\gamma
_{5}.$ In the \textquotedblleft correspondence\textquotedblright\ limit in
which the $\theta $'s become Grassmann variables, the Dirac equation becomes
a constraint imposed on both bosonic ($p$) and fermionic ($\theta ,\theta
_{5}$) variables: 
\begin{equation}
\mathcal{S}_{0}\psi \equiv (p\cdot \theta +m\theta _{5})\psi
=0;\Longrightarrow \mathcal{S}_{0}\equiv (p\cdot \theta +m\theta
_{5})\approx 0.
\end{equation}

For a single free particle, this \textquotedblright classical
Dirac-Equation\textquotedblright\ constraint is supersymmetric\cite{van},
under the supersymmetry generated by $\ p\cdot \theta +\sqrt{-p^{2}}\theta
_{5}$ which however does not leave the position four-vector $x$ invariant .
However, it does leave the \textquotedblleft
zitterbewegungless\textquotedblright\ \ position variable $\tilde{x}^{\mu
}=x^{\mu }+i\theta ^{\mu }\theta _{5}/m$ invariant. \ In the presence of
scalar interaction $M=m+S$ , \ $\tilde{x}^{\mu }$ becomes the
\textquotedblright self-referent\textquotedblright\ form $\tilde{x}^{\mu
}=x^{\mu }+i\theta ^{\mu }\theta _{5}/M(\tilde{x})$ .\ \ \ \ \ The
\textquotedblright pseudoclassical\textquotedblright\ Dirac dynamics is then
governed by the supersymmetric system of constraints 
\begin{equation}
\mathcal{S}=p\cdot \theta +M(\tilde{x})\theta _{5}\approx 0,\ \frac{1}{i}\{%
\mathcal{S},\mathcal{S}\}\equiv \mathcal{H}=p^{2}+M^{2}(\tilde{x})\approx 0.
\label{drc}
\end{equation}%
Since $\theta _{5}^{2}=0$ , the expansion of the self-referent form
truncates so that $M(\tilde{x})=M(x)+i\partial M(x)\cdot \theta \theta
_{5}/M(x)$. Upon quantization, Eqs.(\ref{drc}) then turn into the Dirac
Equation and its standard square when the Grassmann variables become theta
matrices while dynamical variables $x$ and $p$ become their operator
counterparts. Thus the supersymmetry that preserves $\tilde{x}$ is a natural
feature of both the single-particle Dirac equation for the free case and its
standard form for external scalar interaction.

For two particles we introduce interactions that preserve such a
supersymmetry for each spinning particle through the replacement 
\begin{equation}
m_{i}\rightarrow M_{i}(x_{1}-x_{2})\rightarrow M_{i}(\tilde{x}_{1}-\tilde{x}%
_{2})\equiv \tilde{M}_{i},\ i=1,2.
\end{equation}%
The Grassmann Taylor expansions of the $\tilde{M}_{i}$ truncate leading to 
\cite{van} 
\begin{eqnarray}
\mathcal{S}_{1}\psi &=&(\theta _{1}\cdot p+\epsilon _{1}\theta _{1}\cdot 
\hat{P}+M_{1}\theta _{51}-i\partial L\cdot \theta _{2}\theta _{52}\theta
_{51})\psi =0,  \nonumber \\
\mathcal{S}_{2}\psi &=&(-\theta _{2}\cdot p+\epsilon _{2}\theta _{2}\cdot 
\hat{P}+M_{2}\theta _{52}+i\partial L\cdot \theta _{1}\theta _{52}\theta
_{51})\psi =0,  \label{scldr}
\end{eqnarray}%
with the invariants $M_{i}$ and $L(x_{\perp })$ related by Eq.(\ref{ml}).
Eq.(\ref{tbde}) becomes Eq.(\ref{scldr}) when restricted to scalar
interactions. The consequences of pseudoclassical supersymmetries are the
extra spin-dependent recoil corrections to the ordinary one-body Dirac
equations, essential for the compatibility of the two equations.

\subsection{Hyperbolic Form of the Two-Body Dirac Equations for General
Covariant Interactions}

We introduce general interactions by recasting the minimal interaction forms
of the two-body Dirac equations into ones that generalize the hyperbolic
forms we encountered in the treatment of scalar interaction. In the scalar
case, if we begin with the two constraints 
\begin{equation}
\mathbf{S}_{1}\psi \equiv (\mathcal{S}_{10}\cosh (\Delta )+\mathcal{S}%
_{20}\sinh (\Delta ))\psi =0,~\mathbf{S}_{2}\psi \equiv (\mathcal{S}%
_{20}\cosh (\Delta )+\mathcal{S}_{10}\sinh (\Delta ))\psi =0
\end{equation}%
we find that Eqs.(\ref{scldr}) are equivalent to the linear combinations

\begin{equation}
\mathcal{S}_{1}\psi =[\cosh (\Delta )\mathbf{S}_{1}+\sinh (\Delta )\mathbf{S}%
_{2}]\psi =0,~\mathcal{S}_{2}\psi =[\cosh (\Delta )\mathbf{S}_{2}+\sinh
(\Delta )\mathbf{S}_{1}]\psi =0  \label{dcpl}
\end{equation}%
in which the interaction appears through the invariant matrix function $%
\Delta =-\theta _{51}\theta _{52}L(x_{\perp })$ while the $\mathcal{S}%
_{i0}=~(p_{i}\cdot \theta _{i}+m\theta _{5i})$ are free Dirac operators. The 
$\mathbf{S}_{i}$ constraints (and hence the $\mathcal{S}_{i}$ ) are a
compatible pair for general $\Delta $: $[\mathbf{S}_{1},\mathbf{S}_{2}]\psi
=0\ $(and\ $\ [\mathcal{S}_{1},\mathcal{S}_{2}]\psi =0$)$~$provided only
that $\Delta =\Delta (x_{\perp })$ $.$ Consider the four polar and four
axial interactions. \noindent For the polar interactions we find the forms $%
\Delta (x_{\perp })=-L(x_{\perp })\theta _{51}\theta _{52}$ for scalar, $%
J(x_{\perp })\hat{P}\cdot \theta _{1}\hat{P}\cdot \theta _{2}$ and $\mathcal{%
G}(x_{\perp })\theta _{1\perp }\cdot \theta _{2\perp }~$for time-like and
space-like vector respectively and $\Delta (x_{\perp })=\mathcal{F}(x_{\perp
})\theta _{1\perp }\cdot \theta _{2\perp }\theta _{51}\theta _{52}\hat{P}%
\cdot \theta _{1}\hat{P}\cdot \theta _{2}$ for polar tensor. The constraint
equations for vector and scalar interactions given in Eq.(\ref{tbde}) come
from $L$ and the Feynman gauge combination $\mathcal{G}=-J$ $\ $when $\ 
\mathcal{F=}0$. \ The $\mathbf{S}_{i}$ constraints for the axial
counterparts are defined as above but the $\mathcal{S}_{i}$ linear
rearrangements have a minus sign in place of a plus sign. For them the
invariant $\Delta $ forms are $\Delta (x_{\perp })=C(x_{\perp })/2$ for
pseudoscalar , $\Delta (x_{\perp })=H(x_{\perp })\hat{P}\cdot \theta _{1}%
\hat{P}\cdot \theta _{2}\theta _{51}\theta _{52}$ and $I(x_{\perp })\theta
_{1\perp }\cdot \theta _{2\perp }\theta _{51}\theta _{52}\Delta (x_{\perp })~
$for time-like and space-like pseudovector respectively and $\Delta
(x_{\perp })=Y(x_{\perp })\theta _{1\perp }\cdot \theta _{2\perp }\hat{P}%
\cdot \theta _{1}\hat{P}\cdot \theta _{2}~$for axial tensor. Some of these
may find application in meson spectroscopy in addition to the usual scalar
and vector interactions.

\subsection{Strong Potential Forms for Nucleon-Nucleon Scattering}

To represent meson exchanges, we require pseudoscalar interactions as well
as vector and scalar. \ \ Thus we must use

\begin{equation}
\Delta (x_{\perp })=-L(x_{\perp })\theta _{51}\theta _{52}+\mathcal{G}%
(x_{\perp })\theta _{1}\cdot \theta _{2}-C(x_{\perp })/2.
\end{equation}%
with the electromagnetic four-vector (Feynman gauge) condition $J(x_{\perp
})=-\mathcal{G}(x_{\perp })$ relating time and space-like components ($%
\theta _{1}\cdot \theta _{2}=\theta _{1\perp }\cdot \theta _{2\perp }-\hat{P}%
\cdot \theta _{1}\hat{P}\cdot \theta _{2}).$Reduction of the coupled Dirac
equations (\ref{dcpl}) to Schr\"{o}dinger-like form for these combined
interactions produces\cite{bin} for equal masses 
\begin{equation}
\Phi _{w}\rightarrow \Phi _{SI}+\Phi _{D}+(\Phi _{SO}+\Phi _{SOT}\sigma
_{1}\cdot \hat{r}\sigma _{2}\cdot \hat{r})L\cdot (\sigma _{1}+\sigma
_{2})+\Phi _{SS}\sigma _{1}\cdot \sigma _{2}+\Phi _{T}\sigma _{1}\cdot \hat{r%
}\sigma _{2}\cdot \hat{r}.
\end{equation}%
in which \ $L,C,\mathcal{G}$ fix all separate quasipotential pieces. Note
that the quadratic nature of many of the strong potential terms (e.g. $%
S^{2},A^{2},L^{\prime 2},\mathcal{G}^{\prime 2},L^{\prime }C^{\prime }$
etc.) could lead to disastrous results for large coupling-constants.

Beyond the limitation to the above nine mesons, one must choose how the
corresponding nine Yukawa potentials are included in the three invariant
functions $C,L,\mathcal{G}=-J$ . \ We structure our strong potential terms
by assuming that in $\mathcal{G=}$ $-\frac{1}{2}\log (1-\frac{2A}{w})$ we
take $A=\mathcal{A~~}$\ for $\mathcal{A\,}<0$ and 
\begin{eqnarray}
A &=&\frac{w}{\pi }\arctan (\frac{\mathcal{A}\pi }{w})~~,\mathcal{A}>0.\ast
\ast  \nonumber \\
\mathcal{A} &=&g_{\rho }^{2}\vec{\tau}_{1}\cdot \vec{\tau}_{2}\frac{\exp
(-m_{\rho }\bar{r})}{\bar{r}}+g_{w}^{2}\frac{\exp (-m_{\omega }\bar{r})}{%
\bar{r}}+g_{\phi }^{2}\frac{\exp (-m_{\phi }\bar{r})}{\bar{r}}
\end{eqnarray}%
\ \ \ For the invariant $L$ we use Eq.(\ref{ams}) for $S>0$ and 
\[
L=-\frac{1}{2}\log (1-\frac{2S}{w-2A});~~S<0~~~\ast \ast 
\]%
\begin{equation}
\mathcal{S}=-g_{\sigma }^{2}\frac{\exp (-m_{\sigma }\bar{r})}{\bar{r}}%
-g_{f_{0}}^{2}\frac{\exp (-m_{f_{0}}\bar{r})}{\bar{r}}-g_{a_{0}}^{2}\vec{\tau%
}_{1}\cdot \vec{\tau}_{2}\frac{\exp (-m_{a_{0}}\bar{r})}{\bar{r}}.
\end{equation}%
\ 

The modifications (**) of our strong potential terms for large repulsive
vector and large attractive scalar interactions lead to corresponding
changes in the quasipotential portions. For the pseudoscalar invariant
function $C$ we use 
\begin{equation}
C=\frac{1}{w}[g_{\pi }^{2}\vec{\tau}_{1}\cdot \vec{\tau}_{2}\frac{\exp
(-m_{\pi }\bar{r})}{\bar{r}}+g_{\eta }^{2}\frac{\exp (-m_{\eta }\bar{r})}{%
\bar{r}}+g_{\eta ^{\prime }}^{2}\frac{\exp (-m_{\eta ^{\prime }}\bar{r})}{%
\bar{r}}].
\end{equation}
We model effects of form factors by replacing $r$ (=$\sqrt{x_{\perp }^{2}}$)
by $\bar{r}=\sqrt{r^{2}+r_{0}^{2}}~$. In addition we take into account that
the vector mesons may have an anomalous ``magnetic moment''\ type of
coupling. \ The net effect is to include pairs of additional vector and
scalar Yukawa interactions but with opposite signs.

\ For $^{3}S_{1}$ $n-p$ and $^{1}S_{0}$ $n-p$ scattering we obtain excellent
phase shift fits in the energy range from 1 to 350 MeV \cite{bin}. \
Examination of some of the other scattering states shows, however, that the
model needs improvement through inclusion of a) world tensor coupling, b)
pseudovector coupling of the pseudoscalar mesons and c) the off mass shell
effects of the vector meson couplings.

\section{Concluding Remarks}

Nonperturbative solution of our two-body Dirac equations (with all of their
strong-potential structure) in QED has demonstrated that these relativistic
wave equations reproduce the field-theoretic perturbative spectral results
thereby increasing our confidence in their use whenever Coulomb-like
potentials play a significant role in dynamics in field-theoretic or
phenomenological application. They have been successfully applied in QCD
with spectral results as good as those of the most popular approach but
using just two invariant potential functions (with Goldstone boson behavior
as a bonus). Their application to $NN-$scattering necessitates their
extension to include general covariant interactions (achieved through a
natural hyperbolic structure present in them). Phase shift results obtained
from them look promising. In the future, the covariant and local Schr\"{o}%
dinger-like structure of the equations that make them simple to implement
may allow us to combine them with formalisms originally developed for the
nonrelativistic Schr\"{o}dinger equation e.g. the microscopic theory of
meson-meson scattering \cite{barnes} and the unitarized quark model \cite%
{ono}.

\end{document}